\documentclass[twocolumn,showpacs,preprintnumbers,amsmath,amssymb]{revtex4}

\usepackage{graphicx}
\usepackage{dcolumn}
\usepackage{bm}
\usepackage{multirow}
\usepackage{subfigure}

\begin{document}

\title{On the inner Double-Resonance Raman scattering process in bilayer graphene}

\author{D. L. Mafra, E. A. Moujaes, R. W. Nunes and M. A. Pimenta}
\address{Departamento de F\'{\i}sica, Universidade Federal de Minas
Gerais, 30123-970, Belo Horizonte, Brazil.}

\date{\today}

\begin{abstract}

The dispersion of phonons and the electronic structure of graphene
systems can be obtained experimentally from the double-resonance (DR)
Raman features by varying the excitation laser energy. In a previous
resonance Raman investigation of graphene, the electronic structure
was analyzed in the framework of the Slonczewski-Weiss-McClure (SWM)
model, considering the outer DR process. In this work we analyze the
data considering the inner DR process, and obtain SWM parameters that
are in better agreement with those obtained from other experimental
techniques. This result possibly shows that there is still a fundamental open question concerning
the double resonance process in graphene systems.

\end{abstract}

\pacs{63.20.D-, 63.20.kd, 63.22.Rc, 73.22.Pr}
\maketitle

In recent years, the physics of monolayer graphene has been
thoroughly investigated, unveiling a wealth of interesting and
unusual properties, most of which are related to graphene's distinct
electronic properties, that consist of a linear and isotropic
dispersion of the electronic states around the Fermi level ($E_F$)
near the {\bf K} point in the Brillouin zone (BZ). Bilayer graphene
is also a very interesting material. While in the unbiased bilayer
the valence and conduction bands touch each other at the Fermi
level, a gap can be opened and tuned, for example, by the
application of an external electric
field~\cite{gava09,ando09,kuzmenko09,malardprl08,pinczuk09,malardreport},
which makes this a promising system for the fabrication of
nanoelectronic devices. The development of bilayer-graphene-based
bulk devices depends on the detailed understanding of its electronic
properties. Since the unit cell of AB stacked bilayer graphene is
the same as that of graphite, one can model the bilayer electronic
structure using a tight-binding (TB) model for
graphite~\cite{wallace47}, by adapting the Slonczewski-Weiss-McClure
(SWM) parameterization~\cite{McClure1957,SW1958} of relevant
couplings. There are several
theoretical~\cite{mim07,partoens06,gruneis08} and
experimental~\cite{li09,zhang08,mafra09,lmalard07,kuzmenko09,malardphysstatsol}
studies of theses SWM parameters, but the agreement between the
reported values, obtained with different experimental techniques, is
not entirely satisfactory.
\begin{figure}
\includegraphics [scale=0.45]{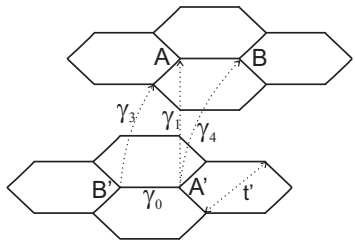}
\caption{\label{bilayer} The intra- ($\gamma_0$ and t$^{\prime}$)
and inter-layer ($\gamma_1$, $\gamma_3$ and $\gamma_4$)
tight-binding parameters in bilayer graphene.}
\end{figure}

In previous resonance Raman studies of bilayer graphene
\cite{lmalard07,mafra09} in our group, the dispersion of the
G$^{\prime}$ Raman band (also called 2D band) as a function of the
laser energy was measured, and the nearest-neighbor hopping parameters
$\gamma_0$, $\gamma_1$, $\gamma_3$ and $\gamma_4$ (shown in
Fig.~\ref{bilayer}) were determined. In Ref.\onlinecite{mafra09}, the
fitting included also the in-plane second-neighbor hopping parameter
$t^{\prime}$, which is expected to be of the same order as the
out-of-plane nearest-neighbor parameters. The parameter $\Delta$,
which represents the difference between the on-site energies of the
sublattices A and B, was also taken into account.

\begin{figure*}
\includegraphics [scale=0.48]{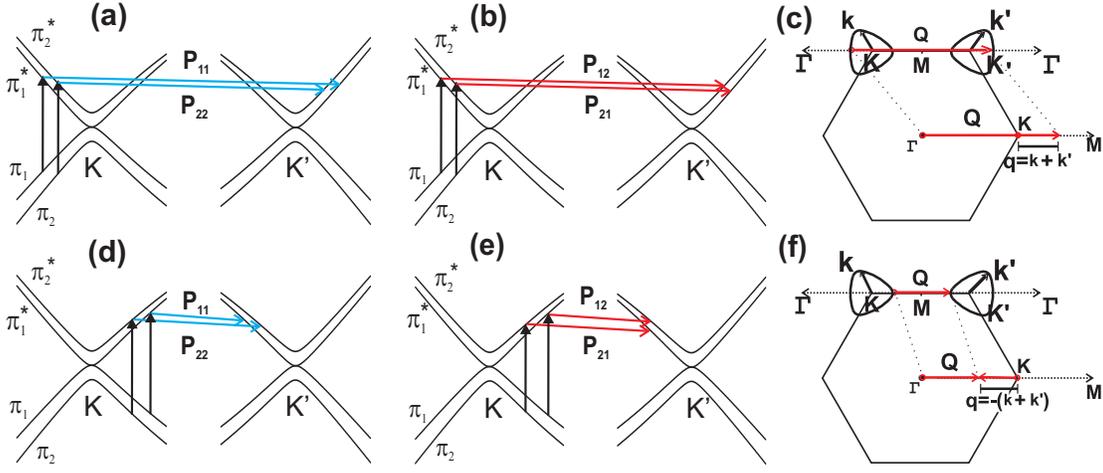}
\caption{\label{processos} (Color online) (a) P$_{11}$ and P$_{22}$ DR
Raman outer processes involving the symmetric phonon. (b) P$_{12}$ and
P$_{21}$ DR Raman outer processes involving the anti-symmetric
phonon. (c) Outer process involving electrons on the {\bf K$\Gamma$}
line and phonons on the {\bf KM} line. (d) P$_{11}$ and P$_{22}$ DR
Raman inner processes involving the symmetric phonon. (e) P$_{12}$ and
P$_{21}$ DR Raman inner processes involving the anti-symmetric
phonon. (f) Inner process involving electrons on the {\bf KM} line and
phonons on the {\bf K$\Gamma$} line.}
\end{figure*}

Group theory analysis for bilayer graphene predicts four distinct DR
processes (P$_{11}$, P$_{22}$, P$_{12}$, and P$_{21}$) along the {\bf
$\Gamma$KM} direction, which are illustrated in
Fig.~\ref{processos}. The triangularly-shaped isoelectronic curves
around the {\bf K} and {\bf K$^{\prime}$} points in
Figs.~\ref{processos}(c) and \ref{processos}(f) are the equienergy
contours of the $\pi$ electrons involved in the scattering
process. The value of the equienergy determined by the laser energy
$E_L$, that creates the electron-hole excitations is the first step of
the DR process, as shown in Figs.~\ref{processos}(a-b) and
\ref{processos}(d-e). The analysis of the data in
Refs.~\onlinecite{lmalard07} and \onlinecite{mafra09} was done
considering a double resonance (DR) process involving only
backscattering of electronic states along the {\bf K$\Gamma$} line
with phonons along the {\bf KM} line, a process we call the outer DR
Raman process. From this analysis, we were able to obtain the SWM
parameters, but our values for the $\gamma_1$ and $\gamma_3$ were
found to be smaller than those determined using other experimental
techniques.

This restricted one-dimensional analysis of the DR process rests on
the following assumptions: (i) that a one-dimensional integration
along the {\bf $\Gamma$KM} direction captures the essential features
of the DR process, as found in Ref.~\onlinecite{maultzsch04b}; (ii)
that some of the graphically determined double-resonant ${\bf q}$
vectors, related to forward-scattering processes [connecting points
on the electronic equienergy curves surrounding points {\bf K} and
{\bf K$^{\prime}$} in Fig.~\ref{processos}(c)], vanish by
destructive interference, as found also in
Ref.~\onlinecite{maultzsch04b}; (iii) that, by plotting the phonon
density of states (PDOS) of graphene satisfying the DR
process~\cite{bob07}, one can identify a strong singularity at the
phonon {\bf q}-vector involved in the outer process, and a much
smaller PDOS value for the phonon {\bf q}-vector along the {\bf
K$\Gamma$} line that backscatters an electronic state along the {\bf
KM} line, and also meets the DR condition, in a process we refer to
as the inner process. This is the reason why attention is usually
paid solely to the one-dimensional outer
process~\cite{bob07,K�rti2002,Ferrari2006}. While the analysis in
Ref.~\onlinecite{maultzsch04b} did consider both outer and inner
processes, and the calculation of the Raman cross section considered
all possible resonant and non-resonant processes, a critical
approximation was employed: the matrix elements involved, related to
electron-photon and electron-phonon couplings, were assumed to be
constant and independent of the wavevectors \textbf{k} and {\bf q}
of the electrons and phonons, respectively.

This state of affairs indicates that some of the conventional wisdom
related to the DR process in graphene-based systems needs to be
reevaluated. In the present work, we address the particular issue of
the preponderance of the outer process. This may be particularly
important in the context of symmetry-breaking potentials imposed on
the graphene system. For example, if a compressive or tensile strain
is applied to the system, either intentionally or due to interaction
with a substrate, one expects the doubly degenerate G$^{\prime}$ peaks
to split into two sub-bands G$_{+}^{\prime }$ and G$_{-}^{\prime}$,
due to the movement of the Dirac cones and consequently the breaking
of the symmetry of the DR Raman process. We reanalyze the data reported in
Ref.~\onlinecite{mafra09} considering now the inner DR process, in
which the electronic excitations involved are along the {\bf KM} line
and the phonons are along the {\bf K$\Gamma$} line. As discussed
below, we obtain a very good fit of the experimental results using the
same values for the parameters $\gamma_0$, $\gamma_4$, $t^\prime$, and
$\Delta$ as in the outer-process case, but with values for $\gamma_1$
and $\gamma_3$ which are in better agreement with those obtained from
other experimental techniques.

In order to understand the experimental results we obtained, it is
important to build a bridge between experiment and theory. To achieve
this, we must find a relation between the electronic and the phonon
dispersions of bilayer graphene. The electronic dispersion of bilayer
graphene can be described in terms of the standard SWM model for
graphite, as mentioned above, using a TB model \cite{wallace47}.
Along the {\bf K$\Gamma$} direction, this amounts to replacing the
parameter $\sigma$ in McClure's expressions \cite{McClure1957} by:
$\sigma=\gamma_0 \left[2\cos\left(2\pi/3 -
\sqrt{3}ka/2\right)+1\right]$.  Here $k$ is the modulus of the
electronic $\bf k$-vector measured from the {\bf K} point and $a =
1.42$~\AA\ is the in plane nearest-neighbor carbon distance. The bands
in the bilayer are obtained from a TB Hamiltonian using the parameters
$\gamma_0$, $\gamma_1$, $\gamma_3$, $\gamma_4$, $t^{\prime}$, and
$\Delta$. Along the high symmetry {\bf K$\Gamma$} direction, the
$4\times4$ matrix factorizes and the dispersion of the four bands are
given by:
\begin{eqnarray}
E_{\pi_2}^{\pi_1^{\ast}}=\frac{-\gamma_1-\sigma
v_3+\Delta^{\prime}\pm\xi_1}{2},\nonumber\\
E_{\pi_1}^{\pi_2^{\ast}}=\frac{\gamma_1+\sigma
v_3+\Delta^{\prime}\pm\xi_2}{2};
\end{eqnarray}
where $v_i=\gamma_i/\gamma_0$~,
\begin{equation*}
\Delta^{\prime} = \Delta +
t^{\prime}\left[2 \cos\left(2\pi/3 - \sqrt{3}ka/2\right)\right.
+\left.\cos\left(4\pi/3 - \sqrt{3}ka \right) \right],
\end{equation*} and
\begin{equation*}
\xi_1^2=\sqrt{(\gamma_1+v_3\sigma\pm\Delta^{\prime})^2+4((1\mp
v_4^2)\sigma^2\mp\sigma v_3 (\Delta^{\prime}\pm\gamma_1))}.
\end{equation*}

It should be noted that these expressions are similar to those found
in Ref.~\cite{lmalard07}, except that here we take $\Delta$ into
account, to allow for the possibility of different doping levels
between the two layers, and also include a $t^\prime$ parameter for
second-nearest-neighbor interactions within the same layer. For any of
the P$_{ij}$ ($i, j = 1, 2$) processes, we seek the dependence of the
phonon energy $E_{ph}$ with $E_{L}$. In the initial step of this
process (electron-hole creation), the incident photon is in resonance
with the excitation of the electronic state from the valence to the
conduction bands at the ${\bf k_i}$ point. In the following, we drop
the vectorial notation for the ${\bf k}$- and ${\bf q}$-vectors, since
we are considering only the {\bf $\Gamma$KM} direction. The laser
energy can then be written as:
\begin{equation}
E_{L}=E_{\pi_i^{\star}}(k_i)-E_{\pi_i}(k_i)\;,
\end{equation}
which allows us to determine the momentum $k_i$ of the excited
electron in the process. The electron is then scattered from a state
in the vicinity of the {\bf K} point to a state in the vicinity of the
{\bf K$^{\prime}$} point by emitting an iTO phonon with energy
\begin{equation}
E_{ph}^{ij}(q_{ij})=E_{\pi_i^{*}}(k_i)- E_{\pi_j^{*}}(k^{\prime}_j)\;;
\end{equation}
where $q_{ij}$ depends on $k_i$ and $k^{\prime}_j$. This equation
uniquely determines the momentum $k^{\prime}_j$ of the scattered
electron, provided that $E_{ph}^{ij}(q_{ij})$ is known. The phonon
energy can be computed, and is directly related to the Raman shift for
a specific P$_{ij}$ process, obtained with a given
$E_{L}$. Physically, the difference between the outer and inner
processes lies in the phonon wavevector $q_{ij}$.

As can be inferred from the geometry in Figs.~\ref{processos}(c) and
\ref{processos}(f), $q_{ij} = k_i+ k^{\prime}_j$ for the outer
process, and $q_{ij} = - (k_i+ k^{\prime}_j$) for the inner process,
measured from the {\bf K} point. In both cases, $q_{ij}$ has a maximum
amplitude of approximately $2k_i$. The outer and inner processes have
their vectors $k_i$ and $k_j^{\prime}$ pointing in opposite
directions. Interestingly enough, this means that calculations for the
inner process can be done simply by switching $k_i$ into $- k_i$ and
$k^{\prime}_j$ into $- k^{\prime}_j$ in Eq.~(1) and looking for values of $k^{\prime}_j$ satisfying Eq.~(3).

\begin{figure}
\includegraphics [scale=0.4]{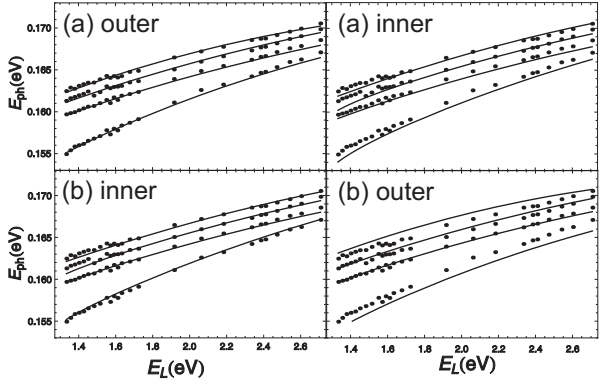}
\caption{\label{fit} Laser energy dependence of the G$^{\prime}$ band
peaks, with four different fittings, considering: ((a) outer) the outer
scattering process with the best SWM parameters for this case, in
Table \ref{tab}(a); ((b) outer) the outer process with SWM parameters of Table
\ref{tab}(b); ((b) inner) the inner scattering process with the best SWM
parameters for this case, shown in Table \ref{tab}(b); ((a) inner) the
inner process with SWM parameters of Table
\ref{tab}(a). Experimental data taken from Ref.\onlinecite{mafra09}.}
\end{figure}

Figure~\ref{fit} shows the experimental data (dots) and the TB fitting
results (curves) for both the outer and the inner processes. Figure
\ref{fit}(a)outer shows the fit considering the outer process for the
DR scattering, i.e., with phonons along the {\bf KM} line. The values
obtained for $\gamma_0$, $\gamma_1$, $\gamma_3$, $\gamma_4$, $\Delta$,
and $t^{\prime}$ are shown in Table \ref{tab}(a). Note that our value
of $\gamma_1=0.35$ eV is slightly smaller than that usually found in
the literature $\gamma_1\approx 0.40$ eV
\cite{zhang08,li09,partoens06}) from other techniques and
calculations. However, the major discrepancy is in the value of
$\gamma_3=0.1$ eV compared to 0.30 eV found in the
graphite literature~\cite{gruneis08,dresselhaus88,schneider09}. Recent
infrared studies in exfoliated bilayer graphene consider
$\gamma_3=0.30$ eV \cite{zhang08}, but this value is not extracted
directly from the experiments.

In order to theoretically compute results and compare with the
experimental data in Fig.~2, we have to consider a form for the phonon
energy $E_{ph}^{ij}(q_{ij})$. In this paper, we consider a nonlinear
relation for the iTO phonon dispersion given by a second-order
polynomial $w(q)=A +Bq+Cq^2$ with $q=k_i+k^{\prime}_j$ and $q=-
(k_i+k^{\prime}_j)$ for the outer and inner processes respectively, as
explained above. We also use two distinct non-linear phonon
dispersions for the symmetric (S) and anti-symmetric (AS)
branches. The best fit was obtained when we consider different
dispersions for the two iTO phonon branches of bilayer graphene. Table
\ref{tab2} shows the parameters obtained for the S and AS phonon
branches, for both scattering processes. Note that the S phonon branch
is more sensitive to the change of scattering process than the AS
mode.
\begin{table}
\centering \caption{Best values of the SWM parameters (in units of
eV) obtained for (a) the outer and (b) the inner scattering
processes.}
 \label{tab}
  \centering
  \begin{tabular}{c|cccccc}
    \hline
    {} & $\gamma_0$ & $\gamma_1$ & $\gamma_3$ & $\gamma_4$ & $\Delta$ & $t^{\prime}$ \\
    \hline\hline\\
    (a) &\,\,\, 3.0\,\,\, &\,\,\, 0.35\,\,\, &\,\,\, 0.10\,\,\, & \,\,\,0.18\,\,\, & \,\,\,0.03\,\,\, & \,\,\,0.10\,\,\, \\
    (b) & 3.0 & 0.40 & 0.30 & 0.18 & 0.03 & 0.10 \\
    \hline\hline
  \end{tabular}
\end{table}

Note also that, in Fig.~\ref{fit}, the fittings considering the two different
DR scattering processes produce rather different results. In the
inner-process case we cannot fit the data with the same parameters
that produce the best fit for the outer process, shown in
Fig.~\ref{fit}(a)outer. This is shown in Fig.~\ref{fit}(a)inner, where
the theoretical curve obtained using the values of $\gamma_1$ and
$\gamma_3$ from Table~I(a) differs considerably from the experimental
values. The best fit for the inner process is shown in
Fig.~\ref{fit}(b)inner, with the values of $\gamma_1=0.40$ eV and
$\gamma_3=0.30$ eV found in the literature, and with the same values
of $\gamma$, $\Delta$ and t$^{\prime}$ as for the outer process. In a
similar manner, the outer process cannot be fitted using the best-fit
parameters for the inner-process case: deviations from the
experimental results are again observed, as shown in
Fig.~\ref{fit}(b)outer.
\begin{table}
\centering \caption{Best values of the iTO phonon dispersion
parameters ($w(q)=A+Bq+Cq^2$) obtained for (a) the outer and (b) the
inner scattering processes.}
 \label{tab2}
  \centering
  \begin{tabular}{c|ccc|ccc}
    \hline
    {} & \multicolumn{3}{c}{symmetric} & \multicolumn{3}{c}{anti-symmetric} \\
    {} &  {A} & {B} & {C} & {A} & {B} & {C} \\
    {} &  (meV) & (meV\AA) & (meV\AA$^2$) & (meV) & (meV\AA) & (meV\AA$^2$) \\
    \hline\hline\\
    (a) & 146.3 & 85.6 & -70.3 & 150.4 & 65.4 & -44.8 \\
    (b) & 150 & -65.9 & -45.6 & 153.4 & -54.9 & -37 \\
    \hline\hline
  \end{tabular}
\end{table}

The above discussion shows that the inner DR process gives TB
parameters in better agreement with those obtained by other
experimental techniques, despite the fact that the number of phonon
states with wavevectors that satisfy the DR process with visible
photons is much larger for the outer DR process \cite{maultzsch04b}.
This result is possibly an evidence that there is still a
fundamental open question concerning the double resonance selection
rules in graphene.

In principle, there is a large number of
phonon wavevectors that satisfy the DR process, connecting electronic
states in equienergy contours around the {\bf K} and {\bf
K$^{\prime}$} points. The calculation of the DR Raman profile must
take into account all these possible DR phonon wavevectors, weighted
by the {\bf q}-vector dependence of the associated electron-phonon
matrix elements. Some interference effects also appear when the Raman
expression is squared out in order to calculate the Raman intensity
\cite{Pimenta2007}. To our knowledge, there is only one full
calculation of the shape of the Raman DR bands that takes into account
the {\bf q}-vector dependence of the electron-phonon matrix elements,
and this calculation predicts an asymmetric shape for the DR bands in
monolayer graphene \cite{Pimenta2007}. However, the experimental DR
Raman bands obtained with visible photons bands are nicely fitted by a
single Lorentzian lines. This possibly shows that some ingredients are
missing in order to fully explain the DR process in graphene systems.

The results obtained in this paper could be the starting point to
investigate other systems which constitute a hot subject in graphene
physics, such as strained or twisted bilayer graphene. In the former
case, bilayer graphene is grown on an insulating material (such as
SiO$_2$), which imposes a strain on the graphene system. This setup
has been considered as a building block for microelectronics
\cite{ni2008}. In the latter case, the stacking of the two layers is
different from the usual AB (Bernal) stacking, being similar to what
is found in naturally occurring and synthetic crystals presenting a
variety of defects, which affect the stacking order mainly in the
c-axis direction \cite{charlier1992,guinea2006}. In both cases, there
is a significant modification in the electronic and optical properties
of the systems, which directly influences the Double Raman bands. This
produces different G$^{\prime}$ peaks that can be theoretically
studied by looking at the P$_{ij}$ processes.

Finally we hope that future experimental work on graphene systems will
reveal more insights on the inner process and on scenarios where it can
be considered more dominant than its counterpart, the outer process. It would also be quite fascinating if one could theoretically
calculate the Raman cross section of bilayer graphene or of any of its
exotic descendants such as twisted or strained graphene systems by
integrating over all k points on the isoenergetic surfaces without any
approximations. This might show that points other than those along the
{\bf K$\Gamma$} and {\bf KM} lines need to be taken into account,
which might lead to a better understanding of the inner and outer
processes.

This work was supported by Rede Nacional de Pesquisa em Nanotubos de
Carbono - MCT, and the Brazilian Agencies CNPq and FAPEMIG.
Resonance Raman studies in the near infrared range were conducted at
the Center for Integrated Nanotechnologies, jointly operated by Los
Alamos and Sandia National Laboratories.

\end{document}